# A thermodynamic unification of jamming

Kevin Lu & H.P. Kavehpour

*Department of Mechanical and Aerospace Engineering, University of California, Los Angeles, Los Angeles, CA 90095, USA*

**Fragile materials[1] ranging from sand to fire-retardant to toothpaste are able to exhibit both solid and fluid-like properties across the jamming transition. Unlike ordinary fusion, systems of grains, foams and colloids jam and cease to flow under conditions that still remain unknown. Here we quantify jamming via a thermodynamic approach by accounting for the structural ageing and the shear-induced compressibility[2] of dry sand. Specifically, the jamming threshold is defined using a non-thermal temperature[3] that measures the 'fluffiness' of a granular mixture. The thermodynamic model, casted in terms of pressure, temperature and free-volume, also successfully predicts the entropic data of five molecular glasses. Notably, the predicted configurational entropy avoids the Kauzmann paradox[4] entirely. Without any free parameters, the proposed equation-of-state also governs the mechanism of shear-banding and the associated features of shear-softening and thickness-invariance[5].**

Despite its mundane appearance, granular materials exhibit a wide range of intriguing phenomena[6,7]. Dry sand, for instance, can deform readily[7] but can also jam abruptly, *e.g.,* the sudden stoppage of flow in an hourglass or a salt-shaker. The abruptness of jamming is commonly quantified by the narrow range of packing-fractions[8] (0.62-0.64) that define the conditions under which the material no longer deforms. Molecular systems also exhibit similar jamming phenomena. For example, liquids such as wood glue become extremely viscous and resistant to flow when cooled within a narrow range of temperatures[9] (2~3°C) below the freezing point. This jamming behavior shared by both granular fluids and viscous liquids is astonishing[6, 10, 11] and suggestive of a common underlying mechanism, but thus-far, a definitive theoretical connection remains unknown.

Jamming was defined[12, 13] as a means to unify all fragile systems[1] and has been qualitatively described using three independent variables: pressure, packing-fraction and an effective temperature[11]. It is known, however, that granular packings are meta-stable: any perturbation in the magnitude or the direction of the applied stress will cause structural ageing[14] where particles rearrange during irreversible compaction.. It is thus erroneous to neglect ageing and assume, for example, that the temperature at which jamming occurs can be defined by pressure and packing-fraction alone. Still, many studies of fragile systems neglect the implications of ageing, possibly because of the narrow range in the temperature and packing density of glassy and granular systems near jamming. Here, we present a new perspective on jamming that includes a connection to the glass-transition of viscous liquids. The proposed equation-of-state (EOS) quantifies jamming as path-dependent states, definable by the stationary observables in pressure, packing-density and shear-rate.

Recent shear flow experiments[2] deduced the EOS of dense granular flows. We observed that the external pressure, $P$, in terms of shear rate, $\dot{\gamma}$, and the free volume[15], $\varepsilon$, has the form

$$P(\varepsilon,\dot{\gamma}) = \frac{1}{\kappa_1} \ln\left[\frac{\varepsilon_o}{\varepsilon} \frac{1}{1 - C\exp(-\kappa_2 \dot{\gamma})}\right]. \tag{1}$$

For dry sand, the constants $\kappa_1 \approx 7\times 10^{-4}$ Pa$^{-1}$ and $\kappa_2 \approx 2\times 10^{-5}$ s. The constant were found to match values found independently from experimental data via the cyclic rule[2]. The free volume $\varepsilon \equiv (V - V_{RCP})$ is the flowing sample volume $V$ referenced to the dynamic random-close-packing volume, $V_{RCP}$. It is normalized by a fit value of the minimum iso-configurational free volume $\varepsilon_0$. As shown in figure 1, equation (1) indicates that the isochoric flows are shear-weakening within intermediate shear-velocities. Spanning five decades from 0.001 to 10 rad s$^{-1}$, pressure 'dips' and reach a minimum between quasi-static and grain-inertial regimes[16]. As one may expect, the weakening mechanism also applies to isobaric flows. Indeed, isobaric shear-compacting is the counterpart to isochoric shear-weakening; the solid volume fraction 'peaks' within intermediate shear-velocities. These isochoric and isobaric flow regimes, however, are interdependent; together, they constitute the transitional regime of granular flow. (A flow sweet-spot is observed near $\dot{\gamma} \approx 200$ s$^{-1}$.)

      The shear-softening scenario presented above has been observed in driven metallic[17] and colloidal[18] glasses, substances which are disordered solids that lack the periodicity of crystals. Why do granular fluids flow like glassy liquids? To explain, we start with the fact that fluid density, enthalpy and viscosity of conventional liquids such as water all scale with the Arrhenius relationship $\sim \exp(H/k_B T)$ where $k_B$ is the Boltzmann constant, $T$ is the temperature, and $H$ is the single 'void-hopping' activation energy[9]. Figure 2 shows that sand compacts during shear, but at a compaction rate that

decreases non-linearly in time with a decaying relaxation constant, $\tau$. The result is fitted using the Kolrausch-Williams-Watts[4, 19] (KWW) equation,

$$\frac{h(t) - h(\infty)}{h(0) - h(\infty)} = \exp\left[-(t/\tau)^\beta\right]. \tag{2}$$

The equation models the normalized change in the column height, $h(t)$, as a function of the relaxation time, $t$, and the Kolrausch exponent, $\beta$. From the fit, we observed that the relaxation is defined by a stretched exponential with a Kolrauch exponent of $\beta \approx 0.6$. This corresponds physically to the multiple relaxation mechanism[4, 9, 15] of granular compaction. It signifies an increase of the apparent activation energy as packing-density increases[19], progressively hindering the process of particle rearrangement. This age-dependent activation energy of sand is an indication of the non-Arrhenius behavior[9, 15] reminiscent of heterogeneous glassy liquids.

The steady-state rheology of figure 1 is ageing or path-independent, based on the reversible branch of packing fraction (0.62-0.64) observed experimentally[8]. The irreversible branch has an expected broader density range (0.555-0.645)[14]. Figure 2, however, suggests that any constitutive model such as equation (1) must account for the compaction of granular flow. This implies that the phenomenological equation (1) must be the solution to a more fundamental theory, one that accounts for the slow structural relaxation even on geological time-scales. To bridge the dynamics of reversible shear flow and irreversible compaction, it is observed that non-thermal grains have a hidden 'ageing' temperature[6], $\Theta \sim 10^{-7}$ J, despite the fact that these particles (size $\gg 1$ μm) are not subjected to thermal fluctuations. To test for its thermodynamic merit, we derive the Helmholtz free energy of flowing sand, $F_{sand}$. From equation (1) using the thermodynamic relation[20] of $P = -(dF/d\varepsilon)$, $F_{sand}$ can be written as

$$\frac{F_{sand}}{N\Theta} = \ln\left(\frac{\varepsilon}{\varepsilon_0}\right) - 1 + \ln\left[1 - C\exp\left(-\frac{\zeta}{\Theta}\right)\right] \qquad (3)$$

where the variables have been recombined into new quantities that are defined as follows: $N \equiv \varepsilon/\nu$ and $\zeta \equiv \kappa_2 \Theta \dot{\gamma}$ where $\nu$ is grain-volume ($\sim 10^{-11}$ m$^3$ for 300 µm particles). Thus far, the manipulation of equation 1 has been strictly algebraic and the original definition of the constants was entirely empiric. However, the recasting suggests specific thermodynamic interpretations of the parameters. The variable $N$ is the number of grains and $\zeta$ is the average dissipation per grain. The test of the model will be an evaluation of how well these interpretations predict behavior in systems other than granular sand.

The free energy of sand makes two critical predictions as confirmed by experiment. First, microscopically, the constant $\kappa_1 = \upsilon/\Theta$ is an elastic property of the material normalized by the only energy scale[6] of the system, $\Theta$. Macroscopically, $\kappa_1$ is deduced from the experiment[2] as $\kappa_1 = -1/\varepsilon (d\varepsilon/dP)_{\dot{\gamma}}$, in a quantity defined as the mechanical compressibility of granular flows. Second, the energy of the flow supplied from the shearing surface is fully dissipated at steady-state. The normalized energy, $\kappa_2 \dot{\gamma}$, would therefore scale as the viscous loss of the flow, $\zeta = \upsilon \eta \dot{\gamma}$ where $\eta$ is the effective viscosity of the granular mixture. Comparing the flow of sand and other fluids drained through a funnel (0.25" opening), we measure a granular viscosity of $\sim 10^{-1}$ Pa-s, which matches that by of mineral oil at room temperature. Altogether, the theoretical values of $\kappa_1 \sim 10^{-4}$ Pa and $\kappa_2 \sim 10^{-5}$ s not only match those fit to the data of figure 1, they also have consistent thermodynamic interpretations.

A unification of jamming must also account for the slow dynamics of glassy systems. From the volume relaxation of figure 2, we observed that sand compacts with a Kolrauch exponent of $\beta \approx 0.6$. Interestingly, typical values of $0.2 < \beta < 1$ are also observed in molecular glasses near jamming. To quantify the dynamics of jamming, we recall that a glass is a liquid in which crystallization is bypassed during cooling[15]. This is the exact scenario exhibited by sand; the angular particles jam because the bulk crystallization never nucleates upon densification. In light of these similarities, we assert that the EOS of equation (1), as a function of the ageing temperature, encompasses the path-dependent states of both jamming and glass-transition. In figure 3, these jammed states are shown by two meta-stable[1] isothermal surfaces, each defined by a particular ageing temperature $\Theta$ used in equation (3).

To substantiate the above claims, Edwards[3] posits that the granular temperature reflects the 'fluffiness' of densely packed grains. To see how 'fluffiness' relates to the particle configuration, we derive the entropy of sand, $S_{sand}$. Using $S = -k_B (dF/d\Theta)$ [20] from equation (3),

$$\Delta S_{sand} = Nk_B - Nk_B \ln(\varepsilon/\varepsilon_0). \qquad (4)$$

Equation (4) accounts for the entropy difference between the jammed and the crystalline states of granular packing, where ideally $\Delta S_{sand} = Nk_B$ is given as the communal entropy[21]. (The communal entropy, $k_B N \approx k\left[\ln(V^N/N!) - \ln(V/N)^N\right]$ via the Stirling approximation[20], accounts for the entropy difference between a liquid and a solid.) In the case for a non-ideal packing (i.e. $\varepsilon > \varepsilon_0$), however, work must be done to constrain the otherwise purely random particles/molecules to sample only the jammed/glassy states[4, 19]—the possible configuration states for all particles. This work reduces the communal entropy by an amount of the configurational entropy, $S^c = Nk_B \ln(\varepsilon/\varepsilon_0)$, scaling in proportional to the volume above ideal packing, $\varepsilon$. In other words, as interpreted from equation (4), equally jammed (or fluffy) configurations can be realized for higher packing densities at the expense of structural order[22].

To verify the configurational entropy $S^c$, we solve equation (1) for $\ln(\varepsilon/\varepsilon_0)$ so that

$$S^c \cong Nk_B \ln\left[1-\exp(-\zeta/\Theta)\right]^{-1}, \quad (5)$$

for $C \approx 1$ and $P \ll k_B T/v$ —which is true for most glasses under atmospheric pressure and thus pressure effects are typically negligible. Figure 4 shows the fit of equation (5) to the configurational entropy data of five different glass-formers. For the fit, the ageing temperature of equation (5) is rescaled as $\Theta = k_B(T - T_0)$, in term of the Kauzmann temperature $T_0$, to preserve the third law of thermodynamics. As it shows, both Kauzmann and fragility plots[4] show good agreement between theory and experiment. More interestingly, the Kauzmann paradox[4, 9]—an unresolved crisis where configurational entropy becomes negative—is entirely avoided.

The shear flow experiment of sand has guided a new classification of jamming as a solid-liquid transition uniquely defined at different structural temperatures. The path-dependent transition is purely kinetic, and yet the transition itself in equilibrium with the structural/ageing temperature. In contrary, other variations[12, 13] rely on an effective granular temperature that is unrelated to the architectural arrangement of particles. Ultimately, the state variables that govern jamming are pressure, shear-rate[17] and the free volume[15].

The EOS for granular flows has provided strong evidence toward the jamming unification of all fragile materials. Broadly speaking, it considers the elastic, the entropic, the free-volume and the hydrodynamic basis of other glass theories presented to-date. This view of jamming applies to phenomena such as stick-slip nucleation in seismic fault ruptures[23], shear-banding in metallic alloys[17], strain-softening in colloidal glasses[18], and even the stop-and-go driving in traffic jams[24]. These types of flows, defiant of conservative fluid models, are closely governed by dynamics that straddle the tipping-point of jamming.

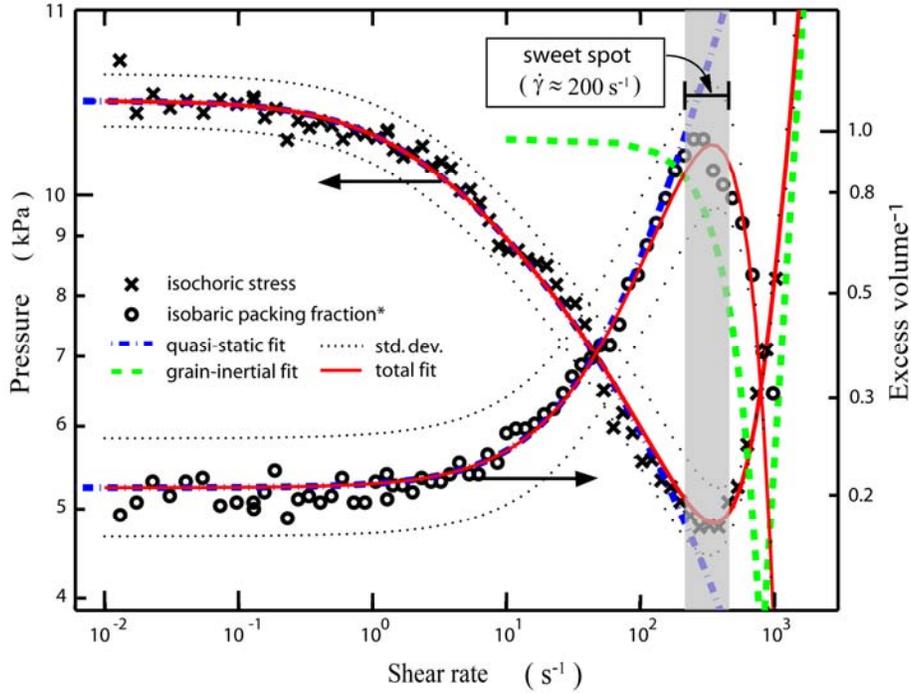

Figure 1 – A log-log plot of both isochoric stress, $\sigma$, and inverse normalized isobaric free volume (i.e. normalized packing fraction), $\varepsilon_0/\varepsilon$, as a function of shear rate, $\dot{\gamma}$. Measurements are made using a torsional rheometer (AR-2000, TA Instruments). Compacted beach sand (grain-size ≈ 438±188 μm, from Santa Monica, CA) is confined concentrically while the top surface shears though logarithmically distributed velocities. The theoretical fit uses equation (1) with an additional grain-inertial term[16], $M\rho D^2 \dot{\gamma}^2$, in terms of grain density, $\rho$, and averaged grain diameter, $D$. The sweet-spot signifies the optimum efficiency in achieving steady-state flow. Shear-rate $\dot{\gamma}$ is calculated based on a two-grain-diameter thickness[2]. The values for the isochoric fit are $C$=0.99±0.004, $\kappa_1$=7.3±0.4×10$^{-4}$ Pa$^{-1}$, $\kappa_2$=2.1±0.8×10$^{-5}$ s, and $M$ = 0.9±0.7×10$^{-3}$. For the isobaric fit, the values are $C$=0.99±0.008, $\kappa_1$=7.0±0.3×10$^{-4}$ Pa$^{-1}$, $\kappa_2$=2.5±0.6×10$^{-5}$ s, and $M$ =2.1±0.5×10$^{-3}$; $\varepsilon_o \approx 4.1 \times 10^{-9}$ m$^3$ from all fits.

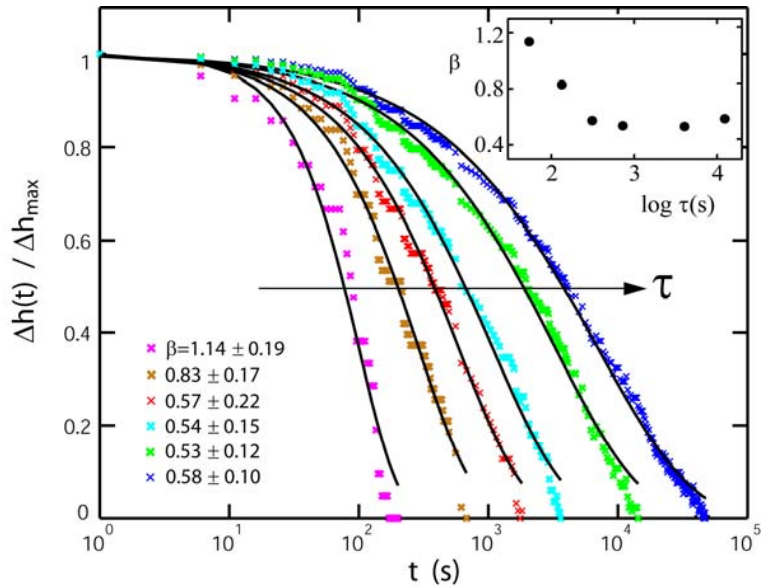

Figure 2 – A semi-log plot of structural aging of sheared granular mixture. The bulk volume is replaced by the total column-height where $\Delta h(t) = h(t) - h(\infty)$ and $\Delta h_{max} = h(0) - h(\infty)$. The structural relaxation, *i.e.* ageing, is non-Arrhenius; different relaxation constants, $\tau$, corresponds to different degrees of structural aging. The relaxation is also non-exponential; the fit constant $\beta$ reaches a steady-state value $\approx 0.5$ after $\approx 1$ hour that reflects the onset of the cooperative rearrangement between grains. The non-Arrhenius and the non-exponential relaxations are reminiscent of the key features of glassy liquids. The sample uses 2.6 g of beach sand sheared at a constant velocity of 0.15 rad s$^{-1}$. The system maintains a constant compression at $\approx 1.5$ kPa while recording height data at 0.1 Hz. The fit uses the KWW relation of equation (2). Note that the entire figure consists of a single experiment where all runs are renormalized by their individual $\Delta h_{max}$.

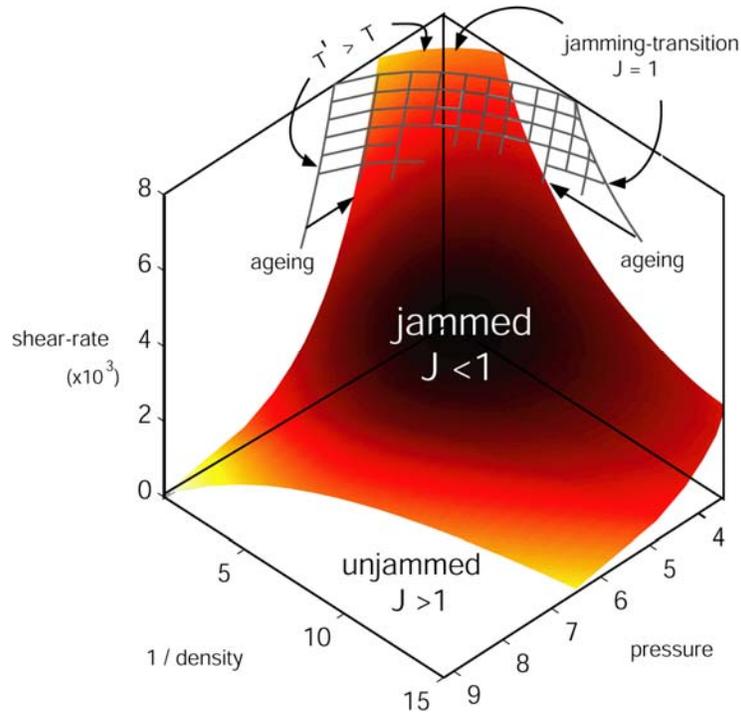

Figure 3 – A plot of the concave jamming-transition surface. The shaded surface corresponds to $J=1$, where $J(T)$ is defined by dividing equation (1) by $\sigma(\varepsilon,\dot{\gamma})$. The state variables are pressure, $\sigma\upsilon/kT$, shear-rate, $\zeta/kT$, and the free-volume, $\varepsilon/\varepsilon_o$, in normalized forms. The jamming transitions given at $J(T)=1$ (solid) and $J(T')=1$ (mesh) signify two equilibrium meta-stable packing arrangements, where the packing at $T$ is denser than the one at $T'$ such that $T'>T$. The unjammed path initiates above the jamming-transition where $J>1$ but terminates at $J=1$ when compaction or structural aging stops within experimental time. This ensures the ageing temperature $T$ remains constant so that the values of $\varepsilon_o$, $\kappa_1$ and $\kappa_2$ are stationary in equation (1). The surface is convex if logarithmic scale were used (see the concavity of the quasi-static fit in Fig. 1).

Nature Physics-2007-09-00990A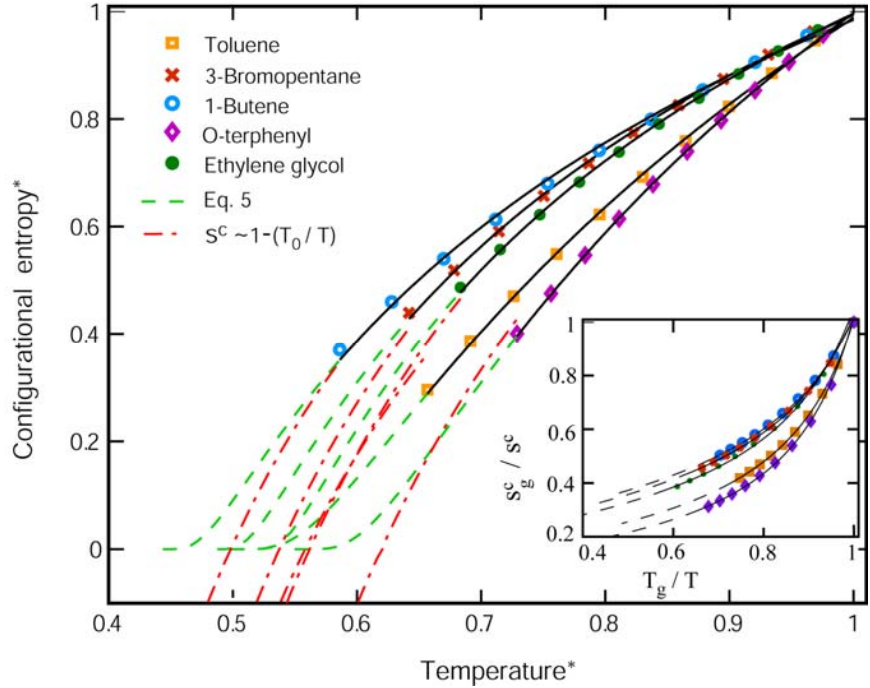

Figure 4 – The Kauzmann plot[4] for five glass-formers of molar configurational entropy*, $s^c/s^c_{fusion}$, versus temperature*, $T/T_{fusion}$, with both quantities normalized by their respective values at the fusion point. The fit agrees well with the entropic data, providing a strong evidence for the unification of granular jamming and glass-transition. Notably, the entropy avoids the Kauzmann paradox (green dash line), in a prediction unlike the one made by the most elegant glass model[4] to-date (red dot-dash line). The inset is the fragility plot of $s^c_{glass}/s^c$ versus $T_{glass}/T$ for the same five glass-formers, and the dash line here is not physical being above the fusion point. The fitting function uses equation (5) where the molar entropy, $s^c$, is given as $s^c(T) \cong xR \ln\left[1 - \exp(-\zeta/k(T-T_0))\right]^{-1}$ for $T \geq T_0$ and $R$ is the universal gas constant. The values of fragility $x$, $\zeta/k$, and the Kauzeman temperature $T_0$, are respectively listed from strong to fragile: 3.24, 8.97 K and 45 K for 1-Butene[25]; 3.55, 17.7 K and 80.6 K (84 K) for 3-Bromopentane[26]; 3.24, 30.7 K and 114 K for Ethylene glycol[27]; 4.22, 38.9 K and 87.8 K (100 K) for Toluene[28]; and 6.01, 65.5 K and 182 K (204 K) for Ortho-terphenyl[29]. The values in parenthesis are the glass-transition temperatures. The matching $x$ between 1-Butane and Ethylene glycol indicates that the fragility index alone can not quantify the glass-transition completely. The molar configurational entropy is derived using the equation,

$$s^c(T) = \Delta_{fusion}s - \int_T^{T_{fusion}} dT' \left[\left(C_p^{liq} - C_p^{cr}\right)/T'\right]$$ for $T \leq T_{fusion}$, and $C_p^{liq}$ and $C_p^{cr}$ are the experimentally measured isobaric (molar) heat-capacities of the liquid and crystalline states. The interpolation/extrapolation of the heat-capacity data were in terms of second-order polynomials.